# Observation of Topologically Robust Localized Magnetic Plasmon Skyrmions

Zi-Lan Deng[1,2,*], Tan Shi[1], Alex Krasnok[2], Xiangping Li[1,*], Andrea Alù[2,3,*]

[1]*Guangdong Provincial Key Laboratory of Optical Fiber Sensing and Communications, Institute of Photonics Technology, Jinan University, Guangzhou 510632, China*

[2]*Photonics Initiative, Advanced Science Research Center, City University of New York, New York, NY 10031, USA*

[3]*Physics Program, Graduate Center, City University of New York, New York, NY 10016, USA*

[*]E-mail: zilandeng@jnu.edu.cn, xiangpingli@jnu.edu.cn, aalu@gc.cuny.edu

**Optical skyrmions have recently been constructed by tailoring electric field distributions through the interference of multiple surface plasmon polaritons, offering promising features for advanced information processing, transport and storage. Here, we construct topologically robust plasmonic skyrmions in a wisely tailored space-coiling meta-structure supporting magnetic-based localized spoof plasmons (LSPs), which are strongly squeezed down to $\lambda^3/10^6$ and do not require stringent external interference conditions. By directly measuring the spatial profile of all three vectorial magnetic fields, we reveal multiple π-twist target skyrmion configurations mapped to multi-resonant near-equidistant LSP eigen-modes. The real-space topological robustness of these skyrmion configurations is confirmed by arbitrary deformations of the meta-structure, demonstrating flexible skyrmionic textures with arbitrary shapes. The observed magnetic LSP skyrmions pave the way to ultra-compact and topologically robust plasmonic devices, such as flexible sensors, wearable electronics and ultra-compact antennas.**

Skyrmions - topologically stable three-dimensional (3D) vector field configurations confined within a two-dimensional (2D) domain - has been prompting significant interest in a number of physical systems, including elementary particles[1], Bose-Einstein condensates[2], nematic liquid crystals[3], and chiral magnets[4]. Beyond elementary skyrmion, nested multiple skyrmions have also been demonstrated in magnetic materials, such as the skyrmionium[5-7] and target skyrmions[8, 9], offering large tunable degrees of freedom in the topological configurations of the skyrmions. As compact and topologically-robust information carriers, skyrmions have been proposed for promising applications such as high-density data storage and transfer[10, 11]. These advances have recently motivated the exploration of optical and plasmonic analogues to skyrmions[12-15]. Unlike topological photonic crystals, where the topological invariants protecting the unusual features of these systems are defined in reciprocal space, optical skyrmions manifest topological properties in real space, offering a topological state of light[12] with promising applications in optical information processing, metrology, transfer and storage. The experimental realization of optical skyrmions has been exclusively demonstrated by interfering or tightly-focusing propagating surface plasmons within a smooth metallic film using carefully controlled external illuminations, and previous optical skyrmion configurations have only been observed in electric or synthetic fields[12-14]. Such interference-based approaches require stringent external excitation conditions, which can only construct regular-shaped single-mode skyrmions at a given frequency.

In this work, we demonstrate localized plasmon skyrmions with topological robustness against arbitrary geometry deformations. Unlike previous optical skyrmion configurations based on propagating surface plasmon wave interference, we realize electromagnetic (EM) skyrmions based on magnetic localized spoof plasmons (LSPs) sustained by a wisely designed space-coiling meta-structure, manifesting additional flexibility and robustness provided by the space-coiling guiding mechanism. The observed LSP magnetic field profile manifests a texture of multiple π-twist concentric features with alternating positive and negative topological charges, which resemble the so-called target skyrmions (TSs)[8, 9]. Our design supports a train of near-equidistant needle-like sharp modes in contrast to conventional localized plasmon resonance spectra with irregular peaks and linewidths[16-19]. Remarkably, the skyrmion number $S$, a topological invariant defined in real space, of the observed skyrmions is unperturbed when the geometry is continuously deformed in arbitrary shapes, leading to topologically robust vectorial field configurations with arbitrary multi-ring profiles, even in the presence of sharp corners and irregular shapes. Such magnetic LSP skyrmions provide a unique way to build arbitrarily shaped skyrmionic textures that is unattainable with previous interference approaches, promising for many flexible and robust applications based on skyrmions.

Our employed meta-structure is composed of a single-armed spiral metallic stripe tightly coiled on itself, with gap width $a$, spiral pitch $d$, radius $R$, thickness $h$, and spiral turn number $n_r$, as shown in the upper panel of Fig. 1. The structure continuous air gap forms a space-coiling region that confines the EM fields at a deeply subwavelength scale. Resonant spectra and field configurations of this space-coiling meta-structure are analyzed using finite element simulations and vectorial near-field measurements in the microwave regime (see Materials and Methods), showing a train of deeply subwavelength resonances at equally-spaced frequencies, $f_0$, $f_0+\Delta f$, $f_0+2\Delta f$, …, where, $f_0$ and $\Delta f$ are the fundamental resonance frequency and free spectral range (FSR), respectively (Middle panel of Fig. 1), determined by the fineness $n_r$ of the space-coiling meta-structure (Supplementary Note 1)[20-22].

Each resonant mode supports a magnetic field profile with axial symmetry, and its unit vector spatially rotates integer multiples of π-twist along the radial direction. As we demonstrate below, the fundamental (π-twist) mode corresponds to an elementary skyrmion, with topological charge 1, and the second (2π-twist) mode forms a skyrmionium[5-7], with topological charge 0. Higher-order modes are multiple-π-twist TSs[8,9], which possess topological charge 1 for odd modes and 0 for even modes due to the accumulated cancellation from adjacent opposite twists. Unlike previous realizations of optical skyrmions, these LSP skyrmion vectorial field profiles correspond to eigenmodes of the space-coiling structure. Hence, they do not require carefully tailored external illuminations and can be excited by various near-field or far-field sources. As the meanderline waveguiding effect is always applicable in our chosen space-coiling geometry, the skyrmion multi-resonances remain stable even when the meta-structure is continuously deformed to arbitrary shapes, as illustrated in the lower panel of Fig. 1.

In order to gain insights into the origin of the scattering response, we first examine a 2D space-coiling cylinder, i.e., metallic stripe spiral of the infinite thickness ($h\rightarrow\infty$), under transverse-magnetic (TM) illumination with magnetic field $H$ pointing along the $z$-direction (Fig. 2a). It has been shown that a textured perfect electric conductor (PEC) surface with multiple disconnected grooves supports LSPs, even if there is no field penetrating the metal[23-25]. On the contrary, our structure is formed by a single connected PEC groove arranged in a space-coiling fashion, supporting purely magnetic LSP modes with only radial lobes (Fig. 2a). Its scattering cross-section (SCS) exhibits a multi-resonant spectrum with equally-spaced needle-like sharp peaks at the deeply subwavelength scale (feature size: R~$\lambda$/1200 at $n_r$=100). This spectrum can be explained by considering the coil as a meanderline waveguide: the resonant peak position and FSR can be predicted by the half-ended effective waveguide model (Supplementary Fig. S1). For feature sizes much smaller than the wavelength, the electric modes do not contribute to the SCS (Supplementary Note 1 and Fig. S2), and the observed LSP resonant features are purely magnetic.

The space-coiling cylinder can be modelled as a homogeneous rod with extremely anisotropic properties (Supplementary Fig. S3), with an anisotropic permittivity tensor with infinite azimuthal- and $z$-oriented components while staying finite in the radial direction. This extreme form of anisotropy supports quasi-static modal distributions, having a radially polarized electric $E$-field and an azimuthal surface current $J$, in contrast to conventional magnetic Mie resonators where $E$ is typically parallel to $J$ [26]. Such exotic resonant mode supports extreme field enhancement (Fig. 2b), as high as $10^5$ in the geometry of Fig. 2. Figure 2c shows the linear dependence of the resonant frequency ($k_0R$) and resonant linewidths ($\Delta k_0R$) with the mode index, confirming its equidistant response. Figure 2d shows that the field confinement factor (resonant wavelength over cavity radius) of the fundamental mode linearly increases with $n_r$, indicating that the resonator compactness can be enhanced by densifying the space-coiling fineness. Figure 2e shows that increasing the meta-structure fineness also increases field enhancement and quality factor, following a quadratic trend. These features are fundamentally limited by the considered finite conductivity of the involved materials, which implies a tradeoff between quality factors and material losses studied in Supplementary Figs. S4a and b. Our study shows that realistic metals can support these resonant responses over a wide frequency range, spanning GHz and THz frequencies, as shown in Supplementary Figs. S4c and d.

Now, we consider finite-thickness space-coiling meta-structures to identify the skyrmion nature of their vectorial field configurations. In the 2D geometry, the LSP field profiles only exhibit scalar (***H***-field) or 2D vectorial (***E***-field) properties. However, for the 3D space-coiling meta-structure with finite thickness, both ***H***-field and ***E***-field distributions manifest 3D vectorial configurations at the interface between the meta-structure and the surrounding background (Supplementary Fig. S5). With decreasing the 3D space-coiling meta-structure thickness, the resonant behavior and the localization of the field pattern are preserved, with a small frequency shift compared to the 2D scenario (Supplementary Fig. S6)[27]. The topological properties of the 3D vectorial field configuration can be quantitatively evaluated by the skyrmion number [12]

$$S = \frac{1}{4\pi} \iint \mathbf{h} \cdot \left( \frac{\partial \mathbf{h}}{\partial x} \times \frac{\partial \mathbf{h}}{\partial y} \right) dxdy, \quad (1)$$

where $\mathbf{h} = \{H_x, H_y, H_z\}/|\mathbf{H}|$ is the local unit vector of the field, and the integrand

$\mathbf{h} \cdot \left( \dfrac{\partial \mathbf{h}}{\partial x} \times \dfrac{\partial \mathbf{h}}{\partial y} \right)$ is the skyrmion density. The skyrmion number is a topological invariant that characterizes the order of topological knots formed by field vectors, i.e., the number of times the field wraps around the unit sphere. The skyrmion number of the magnetic field profile calculated at the air interface (Supplementary Figs. S5a-c and Fig. S6b) is equal to 1, confirming a pure skyrmion field configuration.

To experimentally observe the skyrmion vectorial field configurations, we fabricated an ultra-thin space-coiling meta-structure ($h$=0.016 mm, $a$=1 mm, $d$=1.5 mm, $R$=30 mm, $n_r$=20) over a printed circuit board (Fig. 3a). We scanned the near field of the resonant modes with a 3D scanning platform connected to a vector network analyzer (Fig. 3b). The measured response indeed manifests a multi-resonant spectrum with nearly equidistant sharp peaks (blue curve in Fig. 3c and Supplementary Fig. S18), consistent with the simulation results with the same geometry parameters (red curve in Fig. 3c). Since these modes are eigen-resonances of the meta-structure, we observe a strong excitation of the skyrmions with the near-equidistant multi-resonant spectrum for various near-fields or far-field sources (Supplementary Fig. S7), in stark contrast with previous skyrmions based on the interference of carefully tailored propagating surface plasmons[12-14]. Because these modes resonate at a deeply subwavelength scale with large radial wavevectors $k_r \gg k_0$, the $z$-component wavevector $k_z = -i\sqrt{k_r^2 - k_0^2}$ is mostly imaginary, yielding strong field confinement far beyond the diffraction limit. The measured fundamental skyrmion has a lateral size $d_m = \lambda/100$, and a half vertical size $h_m = \lambda/400$ (Fig. 3d, left-lower panel), leading to an extremely subwavelength mode volume $V_m = 2\pi h_m (d_m/2)^2 = \pi(\lambda/2)^3/10^6$. We stress that such tiny mode volume is obtained with just $n_r$=20, limited by our fabrication and measurement setup. Further squeezing may be achieved by increasing $n_r$. We experimentally observe a relatively high Q-factor of 165 (Fig. 3c), yielding a Purcell factor exceeding $10^7$, promising for various applications requiring strong light-matter interactions. Both out-of-plane (Fig. 3d) and in-plane (Supplementary Fig. S8) magnetic fields along the radial direction have been simulated and measured, yielding excellent agreement, except around the center of the sample, where we observe additional field nulls for the lower modes $m$=1, 2, due to strong coupling between localized EM fields and the magnetic loop probe.

Due to the axial symmetry of the magnetic field profile, the unit vector can be written as $\mathbf{h}(x,y,z) = \{\sin\Theta(\rho)\cos\varphi, \sin\Theta(\rho)\sin\varphi, \cos\Theta(\rho)\}^T$, where $r$ and $\varphi$ are coordinates in the polar system and $\Theta(\rho)$ is the orientation angle of the unit vector. The skyrmion number of the $i^{\text{th}}$ radial mode lobe can be calculated in closed form as

$$S_i = \frac{1}{4\pi} \int_0^{2\pi} d\varphi \int_{\rho_i}^{\rho_{i+1}} d\rho \frac{d\Theta(\rho)}{d\rho} \sin\Theta(\rho) = -\frac{1}{2}\cos\Theta(\rho)\bigg|_{\rho_i}^{\rho_{i+1}}, \qquad (2)$$

showing that the skyrmion number only depends on the initial and final states of $\Theta(\rho)$. Figure 3e shows the unit vectors and $\cos\Theta(\rho)$ distributions along the radial direction. According to Eq. (2), each radial lobe of the mode profile has a skyrmion number +1 or -1, representing an elementary skyrmion polarized in opposite directions. The accumulated total skyrmion number is 1 for odd modes and 0 for even modes, respectively, which realizes a multiple-π-twist TS constructed by multiple elementary skyrmions[8], with rich possibilities to implement various topological configurations of different orders[9].

The vectorial nature of the skyrmion modes can be observed in our real-space measurements of all three magnetic field components, shown in Fig. 4. The in-plane fields are along the radial direction, $H_x$ and $H_y$ reveal a nodal-line profile along their perpendicular axis (*y*-axis and *x*-axis, respectively). At the same time, the out-of-plane component ($H_z$) has only radial variations, in good agreement with our simulations. The mode lobes are distributed purely along the radial direction, in stark contrast to conventional WGM modes with multiple azimuthal modes. The three field components form a hedgehog-like vector configuration (bottom panel of Fig. 4), which is the direct signature of Neel-type skyrmions[28]. This outcome is also confirmed by the skyrmion density and skyrmion number extraction from the field patterns (Supplementary Fig. S11 and S13a).

In contrast to the full skyrmion supported by the ***H***-field, the ***E***-field profile shows a skyrmion configuration with an extra-π/2-twist, yielding a total skyrmion number of 1/2 (Supplementary Fig. S9)[29-31]. The ***H***-field and ***E***-field are parallel to each other due to the extreme anisotropy of the space-coiling structure and have a π/2 relative phase shift (Supplementary Fig. S10, and Movie S1) due to the standing wave nature (along the radial direction) of the resonant mode. Time-varying properties of these skyrmion configurations are shown in Supplementary Movie S1 and S2, indicating that the topological profiles are preserved throughout the entire oscillation period of the EM field, with well-defined skyrmion nature[12].

One unique feature of the skyrmions supported by this spoof plasmon meta-structure is its inherent topological features in real space, which remains robust against continuous geometrical deformations. As the geometry is stretched into an ellipse or deformed to polygonal or even an asymmetric heart shape (Supplementary Fig. S12a), the elementary skyrmion field configuration adapts itself to the new geometrical shape (Supplementary Fig. S12b) without affecting its resonance features. The skyrmion density distribution is modified, with maxima accumulating in regions with sharp curvatures (Supplementary Fig. S12c), but the skyrmion number is strictly preserved, revealing topological robustness. All higher-order skyrmion modes show similar robustness, as their skyrmion number is always 1 for odd modes and 0 for even modes, independent of the geometrical shape (Supplementary Figs. S13 and S14a). Interestingly, the resonance frequencies are weakly affected by these geometrical changes (Supplementary Fig. S14b), realizing in a practical meta-structure with the highly sought property of shape-independent resonance features envisioned in zero-index metamaterials[32].

To experimentally demonstrate this topological robustness, we fabricated space-coiling meta-structures with different shapes: elliptical, polygonal, and heart-shaped, keeping constant the effective coil length. We show the measured spectra in the left-most panels of Figs. 5a-c. Even though the geometries are drastically different, their EM response manifests similar near-equidistant multi-resonant spectra. The spatial profiles of all vectorial magnetic field components of these modes for different shapes are directly measured by our near-field scanning technique (as shown in Supplementary Figs. S15-S17), which agree well with the simulation results, manifesting arbitrarily shaped skyrmion textures indicated by the extracted skyrmion number from the arbitrarily shaped mode profile (Supplementary Fig. S14a). Based on the three field components, we reconstruct their vectorial field configurations in the right panels of Figs. 5a-c, showing multiple nested rings following their shapes, with multiple-π-twists from the geometry center to the periphery. Such arbitrarily shaped skyrmion configurations have never been observed in magnetic materials or optical skyrmions, providing a unique way to realize flexible skyrmions for applications in various technological areas. In addition to continuous shape deformations, we also studied the effect of abrupt defects introduced within the meanderline geometry (Supplementary Note 7 and Fig. S19), particularly in the forms of gaps or shorts in the line. These results show that, as long as the defect does not abruptly modify the field continuity, the structure's skyrmion nature is preserved.

To conclude, in this work we have revealed magnetic LSP skyrmions in a space-coiling meta-structure. The skyrmions stem from the eigen-resonances of the tailored meta-structure, and hence they do not require external interference in the illumination. We have observed multiple-π-twist TS vectorial configurations in real-space through near-field scanning, yielding extremely subwavelength features, down to $\lambda^3/10^6$. The magnetic LSP skyrmions' main feature is its large field profile stability against continuous deformations of the meta-structure geometry due to their topological features together with the broad applicability of the meanderline model for the space-coiling meta-structure, manifesting an overall stable multi-resonant spectrum and flexible skyrmionic textures with arbitrary shape. Although our proof-of-principle has been demonstrated in the microwave regime, we envision exciting opportunities in various frequency ranges, from near-DC to THz regimes. Our findings offer an ideal platform to support the next revolution of information processing with inherent advantages in compactness, stability, and precision for potential applications, including miniaturized spectroscopy, THz sources and microwave photonics.

**Figures**

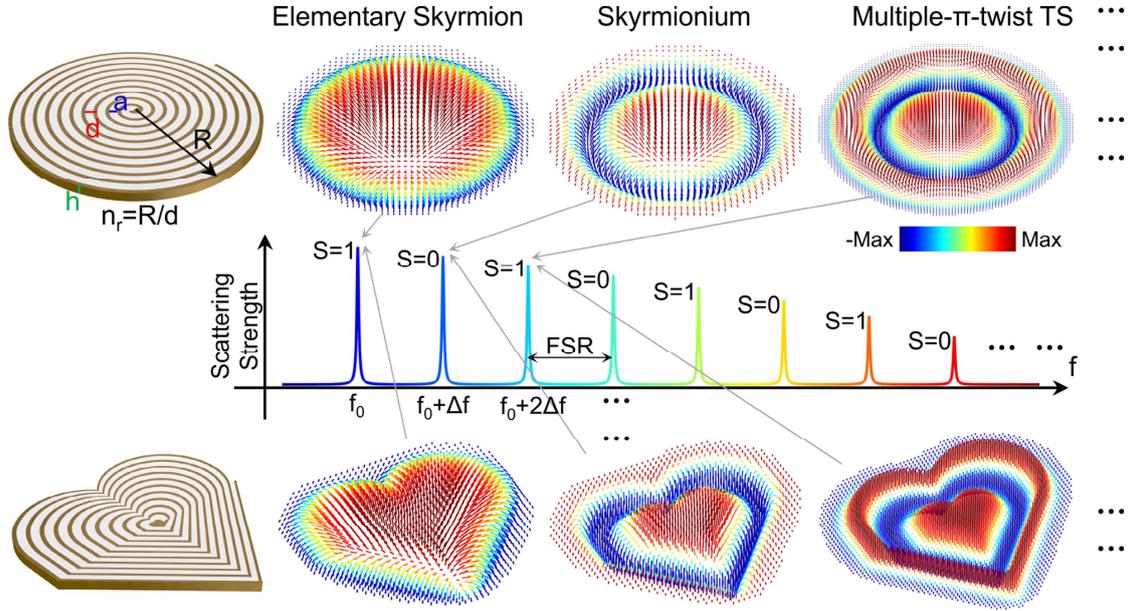

**Figure 1. Topologically robust LSP skyrmions with near-equidistant multi-resonant responses supported by ultrathin space-coiling meta-structures.** The space-coiling meta-structures made of a single-armed metallic spiral strip with pitch size *d*, gap width *a*, radius *R* and number of turns $n_r$ are shown on the left, the vectorial magnetic field configurations of their supported modes are shown on the right. The space-coiling meta-structures support a train of deeply subwavelength LSP modes with near-equidistant multi-resonant response, whose vectorial magnetic field configurations are characterized by elementary skyrmion, skyrmionium and multiple-π-twist TS, with fundamental resonance frequency $f_0$, FSR $\Delta f$, and skyrmion number *S*=1 for odd modes and S=0 for even modes (middle panel). These topological features are robust against continuous deformations of the geometry, for instance from round (upper panel) to asymmetric heart shape (lower panel).

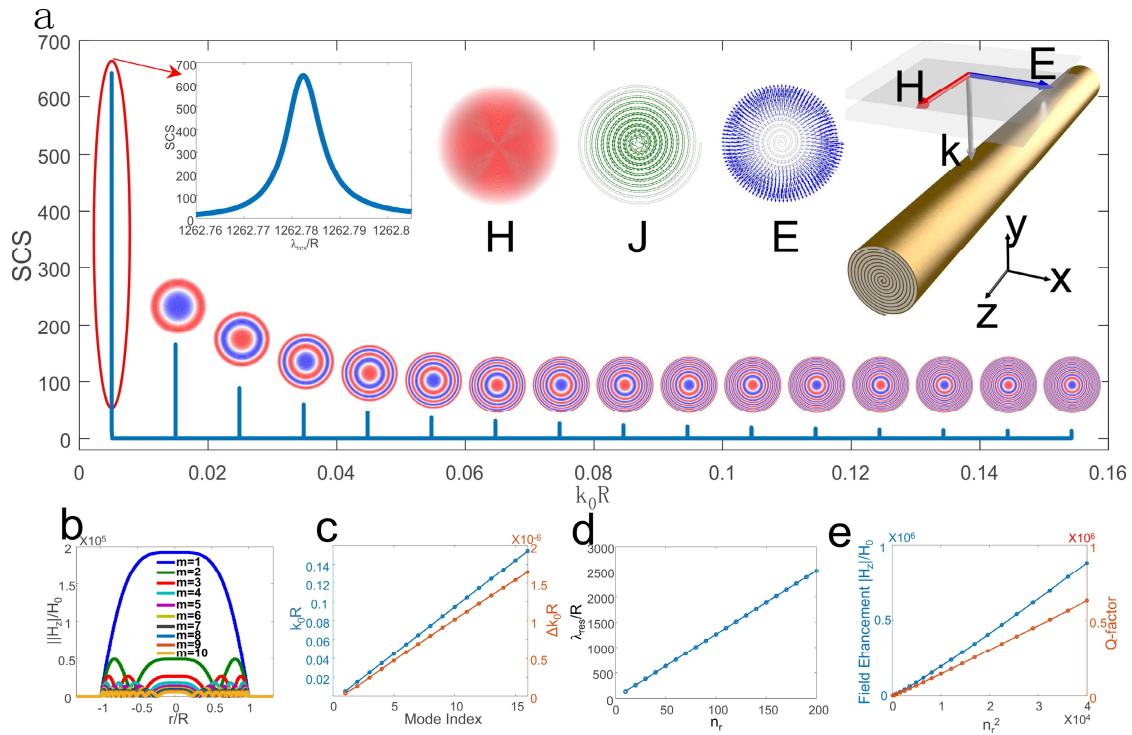

**Figure 2. 2D spoof magnetic plasmons with near-equidistant multi-resonant response supported within a deeply subwavelength space-coiling cylinder.** (a) Scattering spectrum of a PEC space-coiling cylinder with spiral turns $n_r$=100, duty cycle $a/d$=2/3, illuminated by a TM plane wave. The upper insets show the enlarged scattering spectrum, magnetic field (**H**), surface current (**J**), and electric field (**E**) patterns of the fundamental mode, and schematic of the structure. The lower insets show the magnetic field patterns for higher-order resonance modes. (b) Magnetic field enhancement of the resonant modes along the radial direction, normalized to the impinging field. (c) Normalized resonant frequency $k_0R$ of the first 16 modes. (d) Resonant wavelength of the fundamental mode as a function of the spiral turn number $n_r$. (e) Field enhancement and quality factor as a function of $n_r^2$.
16

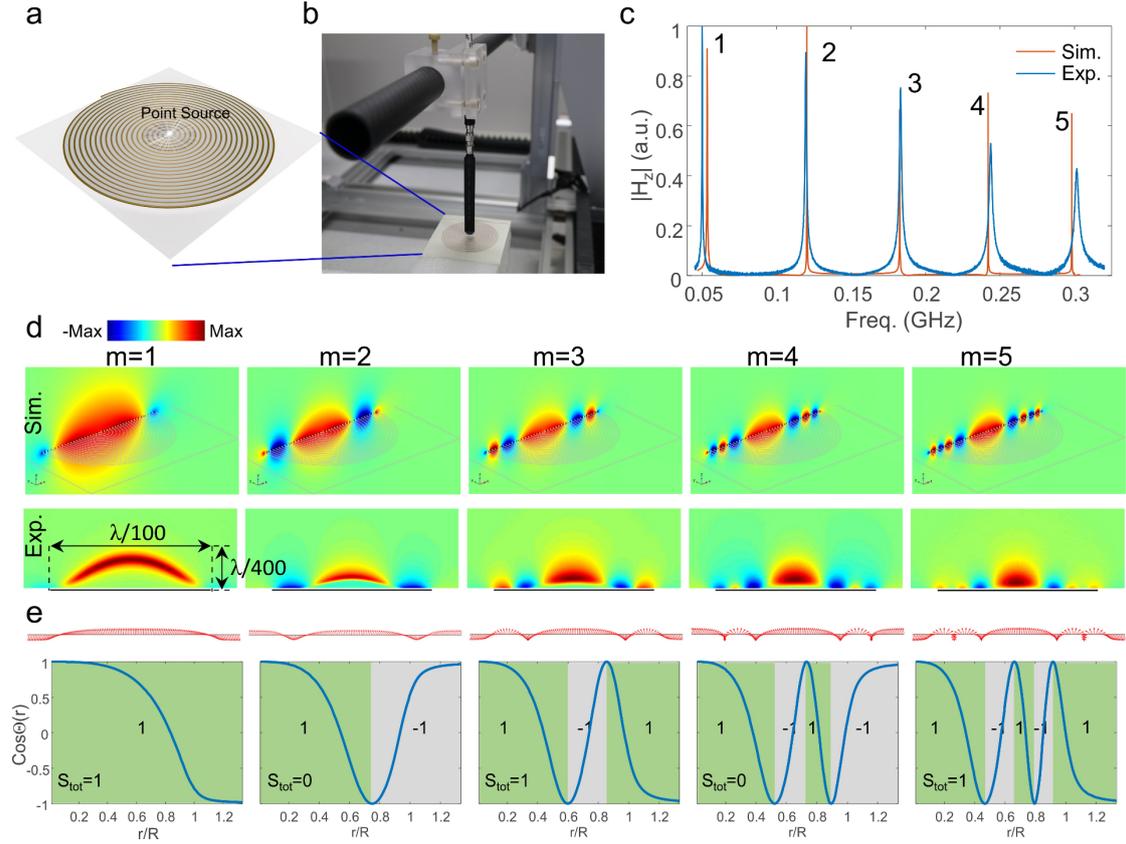

**Figure 3. Magnetic field nature and topological features of LSP skyrmions in an ultrathin space-coiling meta-structure.** (a) Schematic of the fabricated space-coiling meta-structure. (b) Experimental setup for near-field scanning of all vectorial magnetic field components of the skyrmion mode. (c) Measured (blue) and simulated (red) near-equidistant multi-resonant response spectra containing the first five modes. Measured quality factors for those modes are 165, 78, 62, 45, and 34, respectively. (d) Simulated (upper panel) and measured (lower panel) magnetic field ($H_z$) profile at the central plane perpendicular to the sample (*xz*-plane). (e) Unit vector configuration of magnetic field distribution (upper panel) and cosine of the unit field orientation angle distribution (lower panel) along the radial direction for each mode. Insets in the lower panel show the skyrmion number in each mode lobe, as well as the total skyrmion number.

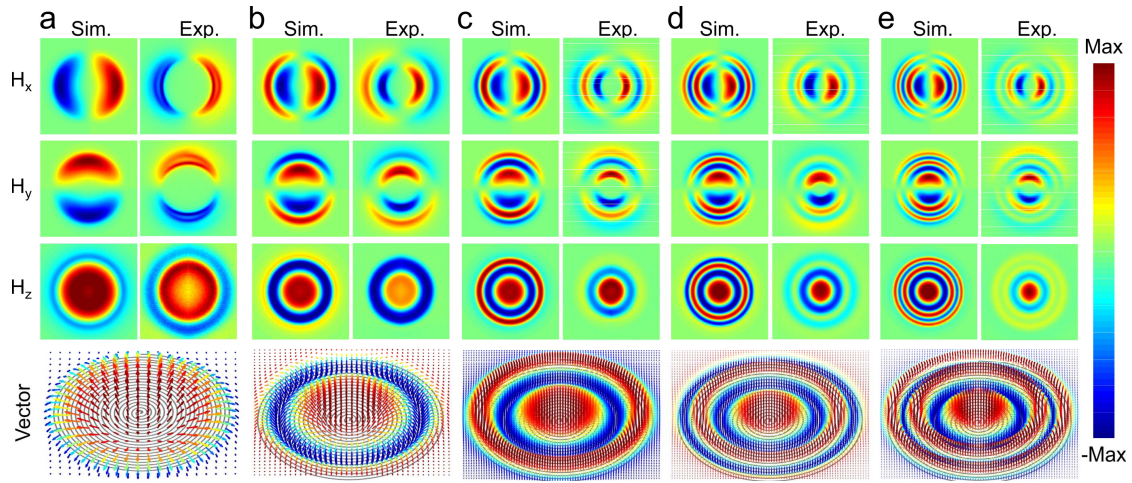

**Figure 4. Real-space imaging of the vectorial magnetic skyrmion configurations.** (a-e) Upper panels: Simulated and measured 3D vectorial components of magnetic field ($H_x$, $H_y$, $H_z$) patterns; Bottom panel: unit vector field configurations of the EM skyrmion modes, of mode index (a) m=1, (b) m=2, (c) m=3, (d) m=4, and (e) m=5, respectively.

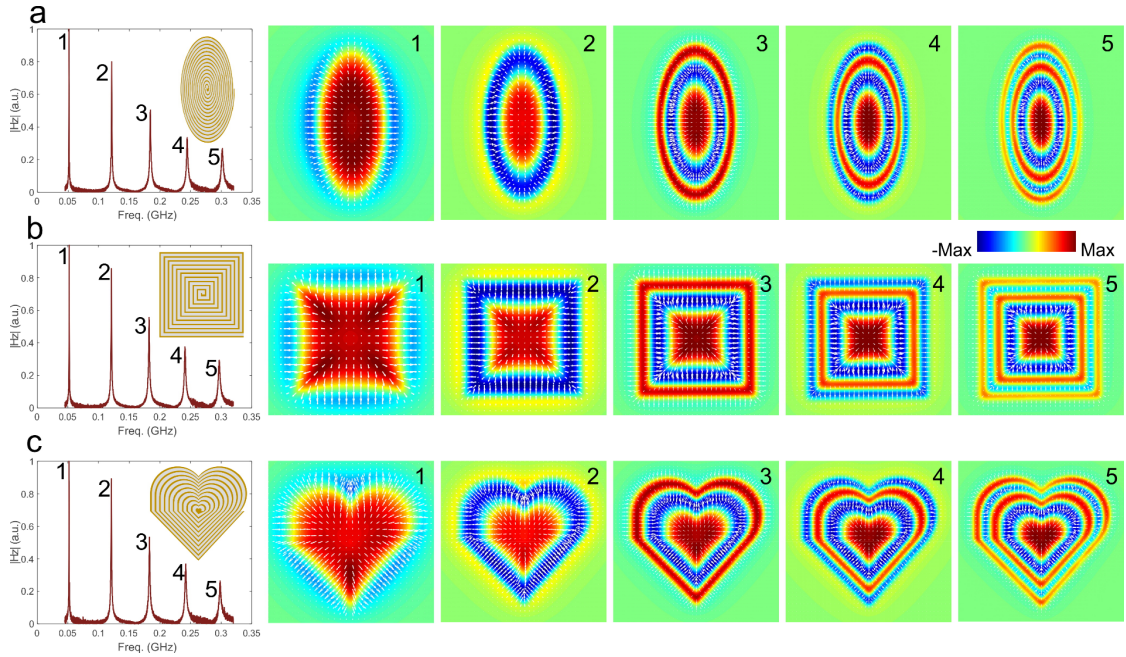

**Figure 5. LSP skyrmions with robust properties against continuous shape deformations of the space-coiling meta-structure.** Measured near-equidistant multi-resonant response spectra, geometry schematics (left-most panels) and vector field configurations with out-of-plane components illustrated by colours and in-plane components by arrows (right 5 panels) for different geometries: (a) ellipse, (b) square, (c) heart-shape.

19